# Highly sensitive and high throughput magnetic resonance thermometry of fluids using superparamagnetic nanoparticles

*Darshan Chalise [1,2*], David G. Cahill[1,2,3]*

[1]*Department of Physics, University of Illinois at Urbana-Champaign, Urbana, IL, 61801, USA*

[2]*Materials Research Laboratory, University of Illinois at Urbana-Champaign, Urbana, IL, 61801, USA*

[3]*Materials Science and Engineering, University of Illinois at Urbana-Champaign, Urbana, IL, 61801, USA*

*Corresponding Author – darshan2@illinois.edu

**Abstract**

Magnetic resonance imaging (MRI) enables non-invasive 3D thermometry which has potential applications in biological tissues and engineering systems. In biological tissues, where MRI is routinely used to monitor temperature during thermal therapies, $T_1$ or $T_2$ contrast in water are relatively insensitive to temperature, and techniques with greater temperature sensitivity such as chemical shift or diffusion imaging suffer from motional artifacts and long scan times. MR thermometry is not well developed for non-biological or engineering systems. We describe an approach for highly sensitive and high throughput MR thermometry that is not susceptible to motional artifacts and could be applied to various biological systems and engineering fluids. We use superparamagnetic iron oxide nanoparticles (SPIONs) to spoil $T_2$ of water protons. Motional narrowing results in proportionality between $T_2$ and the diffusion constant, dependent only on the temperature in a specific environment. Our results show, for pure water, the nuclear magnetic resonance (NMR) linewidth and $T_2$ follow the same temperature dependence as the self-diffusion constant of water. Thus, a $T_2$ mapping is a diffusion mapping in the presence of SPIONs, and $T_2$ is a thermometer. For pure water, a $T_2$ mapping of a 64 × 64 image (voxel size = 0.5 mm × 0.5 mm × 3 mm) in a 9.4 T MRI scanner resulted in a temperature resolution of 0.5 K for a scan time of 2 minutes. This indicates a highly sensitive and high throughput MR thermometry technique that potentially has a range of applications from thermal management fluids to biological tissues.

## I. INTRODUCTION

Temperature monitoring in biological systems is important during thermal ablation of cancerous tissues to ensure effective treatment and prevent collateral damage of healthy tissues [1]. Temperature is also an indicator of physiological processes [2], and therefore, accurate determination of temperature can be used for diagnosis of diseases. In engineering systems, temperature monitoring of heat sinks or heat transfer fluids would be important to ensure effective thermal management of devices [3].

Due to its ability to obtain non-invasive 3D images, MRI has been routinely used for temperature monitoring during thermal therapies. Current MRI thermometry in biological systems often utilizes the temperature dependence of the spin-lattice relaxation time ($T_1$) of water protons [1]. $T_1$ is sensitive to temperature as it depends on the diffusion [4]. While the translational diffusion in water is highly sensitive to temperature (a 1% change in temperature corresponds to a 4% change in the diffusion constant at room temperature), the rotational diffusion is not sensitive to temperature [4]. This results in lowering the overall sensitivity of $T_1$ to temperature. For example, in pure water, a 1% change in temperature typically corresponds to a 3% change in $T_1$ [1,4]. Due to intrinsically long acquisition times used in $T_1$ sequences (inversion recovery or saturation recovery), a reasonable acquisition time often results in a temperature resolution of >1.5 K [5]. In many biological systems, due to fast diffusion, water $T_2$ is often limited by $T_1$ and therefore, $T_2$ imaging is also not a sensitive thermometer [1].

In water, highly sensitive MR temperature imaging is performed with proton resonance frequency (PRF) shift imaging, also known as chemical shift imaging (CSI) [6]. The chemical shift of protons changes by ~0.01 ppm/K for water in most tissues due to the strong temperature dependence of hydrogen bonding; since the linewidth of water protons is very narrow, a temperature resolution of ~ 0.5 K can be achieved with chemical shift imaging [6]. However, chemical shift imaging is extremely time consuming as the image is entirely phase encoded [1]; an $n \times n$ image slice typically takes $n$ times greater scan time than $T_1$ or $T_2$ imaging. More importantly, the need to obtain a baseline image for phase contrast means that motion in between two images causes significant motional artifacts [1].

In fluids where hydrogen bonding is not present, a strong temperature dependence of chemical shift is not expected [1]. Thus, CSI would not be a sensitive thermometry in thermal management fluids other than water.

For pure water, near room temperature, the translational diffusion ($D$) is also highly temperature dependent due to the strong temperature dependence of hydrogen bonding in water [6]. The sensitivity coefficient, which describes the percentage change in the diffusion constant created by a one percent change in temperature is approximately 4 [7]. The sensitivity coefficient is similar for water protons in different tissues [8–10].

The translational diffusion constants of mineral oils with large molecules, which could be used as dielectric fluids for cooling applications, also show a strong temperature dependence [11]. Engineered dielectric fluids used for heat transfer applications also show a strong temperature dependence of kinematic viscosity [12] and can be expected to have a strong temperature dependence on diffusion. Therefore, the translational diffusion constant can be a sensitive thermometer in both biological systems and in thermal management fluids.

Diffusion mapping, however, requires long acquisition times and suffers from interscan gradient instabilities [4]. More significantly, similar to CSI, a diffusion mapped MRI image is susceptible to interscan motional artifacts [4]. This restricts the application of diffusion mapping in living organisms as well as in fluid cooled engineering systems where fluid motion enhances heat transfer.

Apart from MR thermometry using intrinsic properties such as relaxation times, diffusion constant, or chemical shift, various thermometry techniques have been suggested utilizing MR contrast agents. Among these, phase change contrast agents have been proposed as thermometers [1]. However, phase change contrast agents only provide a semi-quantitative means of thermometry. Some recent thermometry techniques utilizing contrast agents includes contrast agents that induce large temperature dependence of chemical shifts (1 ppm/K) [13] or field inhomogeneities (or equivalently $T_2^*$) [14]. While CSI is limited in practical application due to long scan times and motional artifacts, $T_2^*$ imaging utilizing the temperature dependence of magnetization is limited in temperature range and is susceptible to external field inhomogeneities.

Superparamagnetic nanoparticles of $Fe_3O_4$ are routinely used in biological systems for targeted drug delivery [15], MR $T_1$/$T_2$ contrast [16], and magnetic particle imaging [17]. The synthesis and biomedical applications of superparamagnetic $Fe_3O_4$ is reviewed in Ref. [18]. Here, we demonstrate a highly sensitive method utilizing $Fe_3O_4$ nanoparticles as $T_2$ contrast agents, in which we image diffusion, and therefore temperature, as a form of $T_2$ contrast.

$T_2$ contrast imaging is significantly faster than conventional diffusion or PRF shift imaging [1]. More importantly, $T_2$ contrast or mapping does not require a baseline image [1], making this method of thermometry less susceptible to interscan motional artifacts compared to diffusion or chemical shift mapping.

SPIONs show almost no cytotoxicity in most cells up to a SPION volume fraction of 20 ppm [19]. The diffusion constant of water in tissues [8,10] and the self-diffusion constant of dielectric fluids have strong temperature dependence. Therefore, this thermometry technique is potentially utilizable in several biological and engineering applications.

## II. THEORY

The strong magnetization of the superparamagnetic particles introduces field inhomogeneities and increases the transverse relaxation rate, $R_2$, in a liquid [20–23], i.e.

$$R_2 = R_{2,0} + R_{2,S}. \qquad (1)$$

Here, $R_{2,0}$ is the relaxation rate in a liquid in the absence of the SPIONs and $R_{2,S}$ is the relaxation rate due to the SPIONs.

As long as the correlation time of protons to diffuse past the SPION ($\tau_c = \frac{d^2}{4D}$, where, $d$ is the diameter of a SPION and $D$ is the translational diffusion constant of protons in the liquid) is smaller than the reciprocal of the resonance frequency shift corresponding to the field inhomogeneity induced by the SPOIN, i.e., $\Delta\omega\tau_c \leq 1$, motional narrowing occurs. When motional narrowing takes place, the relaxation induced by SPIONs can be estimated by outer sphere theory of the relaxation of protons in the presence of paramagnetic impurities [20–23]:

$$R_{2,S} = \frac{1}{T_2^*} = \frac{5}{9} v_f \tau_c (\Delta\omega)^2. \qquad (2)$$

Here, $v_f$ is the volume fraction of the SPIONs in the liquid and $\tau_c$ is the correlation time of the proton, $\Delta\omega$ is the average frequency shift of a proton adjacent to the surface of the SPION and is given by:

$$\Delta\omega = \sqrt{\frac{4}{5}} \frac{\gamma \mu_0 M_p}{3}. \tag{3}$$

Here, $\gamma = 267.52$ Mrad/T is the gyromagnetic ration of the proton, $\mu_0$ is the permeability of free space and $M_p$ is the magnetization of an individual SPION. Our NMR measurements (see the Results section below) show that the outer sphere theory describes our data extremely well (within a factor of 1.2), and Eq. 2 is a reliable estimate of the relaxation rate due to SPIONs.

If the relaxation rate induced by SPIONs is much greater than the natural relaxation rate of a proton in a liquid, i.e., $R_2 = R_{2,0} + R_{2,SPION} \approx R_{2,SPION}$, and a $T_2$ experiment satisfies $\tau_c \ll T_{echo}$, echo pulses are unable to recover the relaxation of protons due to SPIONs [21], and

$$\frac{1}{T_2} = \frac{1}{T_2^*} = R_{2,S} = av_f \tau_c (\Delta\omega)^2 = av_f \frac{d^2}{4D} \left(\frac{\frac{\gamma}{2\pi}\mu_0 M_p}{3}\right)^2. \tag{4}$$

Therefore, in the motional narrowing regime, if the magnetization of the SPIONs remain constant, and $T_{echo} \gg \tau_c$, proton $T_2$ in the presence of SPIONs scales linearly with the diffusion constant $D$, and $T_2$ contrast imaging is essentially equivalent to diffusion contrast imaging.

In using the outer sphere theory, we have neglected the overlap of magnetic fields induced by individual SPIONs. Even for a volume of SPIONs as high as 1000 ppm, the average distance between the nanoparticles is ~ 120 nm. Since the magnetic field due to dipole decreases as $1/r^3$ [23], the average field experienced by one SPION due to another SPION is less than 0.1% of the average field experienced by water molecules. Therefore, the magnetic interaction between the SPIONs can be neglected.

From the outer sphere theory, the temperature dependence of the relaxation time comes from the temperature dependence of the self-diffusion constant and the temperature dependence of the magnetization of the nanoparticles. Near room temperature, a 1% change in absolute temperature results in approximately a 4% change in the self-diffusion constant of water [7]; therefore, the sensitivity of the change in diffusion constant is 4. For $Fe_3O_4$ nanoparticles, the change in magnetization with temperature is small until the temperature approaches the Curie temperature (858 K for bulk $Fe_3O_4$ and ~850 K for 15 nm particles [24]). For $Fe_3O_4$ with diameter ~15 nm, the magnetization decreases approximately 6% when the temperature increases by 27%, from 278 K (5° C) to 353 K (80 ° C) [25]. Thus, the sensitivity of the change in magnetization to temperature is ~ 0.2. Since both the increase in the self-diffusion constant and the decrease in magnetization cause an increase in the relaxation time ($T_2$), the sensitivity of our method in pure water is ~ 4.2.

## III. METHODS

**SQUID Magnetometry**

We used commercially available $Fe_3O_4$ nanoparticles in our experiments. $Fe_3O_4$ nanoparticles suspended in water were purchased from Sigma Aldrich (Product: 900043-5ML). Sigma Aldrich specified the concentration of particles in water to be 5 mg/ml using inductively coupled plasma (ICP) analysis. SQUID magnetometry was performed to confirm superparamagnetic behavior and to estimate the magnetic moment and magnetization of the nanoparticles.

SQUID magnetometry was performed on a Quantum Design MPMS 3 at 300 K with a field sweep from -0.3 T to 0.3 T. Colloidal suspension of iron-oxide nanoparticles (0.18 ml) was held using a liquid sample holder (QD 8505-013).

**NMR measurements**

NMR linewidth and $T_2$ measurements were performed using a Varian Unity/Inova system using a 14.1 T (600 MHz) field and a 5 mm broadband probe. Free induction decay (FID) spectra were collected at different temperatures using a relaxation delay ($d_1$) of 4 s, an acquisition time of 1s, a spectral width of 15 kHz, and with 4 signal averages. Shimming the magnetic field was not possible for samples with the superparamagnetic particles, and the shim conditions for pure water were used for samples with the particles. A CPMG pulse sequence [26] was used for determination of $T_2$ with a pulse spacing of 100 $\mu$s with all other acquisition parameters identical to the FID measurements.

Sample temperature in the NMR probe were calibrated using the spectra of ethylene glycol above 35° C and using methanol from 5° C to 35° C at specific temperature set points controlled by an FTS temperature control system [27].

Linewidth and $T_2$ values were determined after data processing in Mnova.

**MRI measurements**

MRI measurements were performed using a Bruker BioSpec 9.4 T preclinical MRI system using an 80 mm volume coil. Isothermal $T_2$ and $T_2^*$ measurements were performed on a 3 mm coronal slice (along the cross section of a cylindrical sample holder) with an image size of 64 × 64 pixels and a voxel size of 0.5 mm × 0.5 mm (× 3mm). $T_2$ mapping was obtained using a multispin multisecho (MSME) sequence [28] with an echo spacing of 5 ms and a total of 12 echo images. $T_2^*$ mapping was performed using multi-gradient echo $T_2^*$ map sequence with gradient echo spacing of 5 ms. A relaxation delay of 2 s was used for all acquisition. The signal was averaged two times. Both $T_2$ mapping $T_2^*$ mapping resulted in a total acquisition time of 2 minutes and 20 seconds.

Isothermal heating of the samples was performed using a water pillow with heated water supplied using a Thermo Scientific Precision water bath. The isothermal condition of a MR slice was confirmed using two type-T thermocouples adjacent to the slice.

Image intensity analysis for $T_2$ and $T_2^*$ mapping, as well as measurement of average variation in $T_2$ and $T_2^*$, were performed using Matlab.

## IV. RESULTS

**Characterization of superparamagnetic iron oxide nanoparticles using SQUID magnetometry**

SQUID measurements were performed at 300 K with a field sweep of -0.3 T to 0.3 T (See Fig S1 in the Supplemental Materials [29]). The results are consistent with the superparamagnetic behavior of the iron oxide nanoparticles [30], as no hysteresis is observed. The superparamagnetic behavior assures there is no aggregation of the particles.

The results of the SQUID magnetometry are fit with a Langevin function:

$$M_{SQ} = M_p \left( \coth\left(\frac{\mu_p B}{k_B T}\right) - \frac{k_B T}{\mu_p B} \right) = \frac{\mu_p}{V_p} \left( \coth\left(\frac{\mu_p B}{k_B T}\right) - \frac{k_B T}{\mu_p B} \right) \quad (5)$$

Here, $M_{SQ}$ is the magnetization measured by SQUID magnetometry, $M_p$ is the magnetization of the particle with magnetic moment $\mu_p$ and volume $V_p$. Based on the slope of the Langevin function, $\mu_p$ is determined to be ~ $3.7 \times 10^{-19}$ Nm/T. $M_{SQ}$ is determined to be 59 emu/gm (~ $3.04 \times 10^5$ A/m) from the magnitude of the Langevin function. This value of saturation magnetization is similar to values reported in literature [31]. The estimate assumes the density of nanoparticles to be 5.17 gm/cm$^3$.

Since the saturation magnetization from the SQUID measurement equals the magnetization of the individual particle, and the magnetic moment is also determined from SQUID measurements, we can determine the average particle volume and, therefore, the volume weighted average diameter of the particles. Based on the SQUID measurement, the average particle volume is ~ $1.3 \times 10^{-24}$ $m^3$ and the volume weighted average particle diameter is ~ 13.5 nm. Thus, the SQUID measurement has allowed us to refine the estimate of the particle diameter of 15 nm ± 2 nm provided by Sigma Aldrich using transmission electron microscopy.

**NMR linewidth and T$_2$ measurements**

NMR measurements were performed on a 14.1 T Varian Unity/Inova NMR spectrometer to confirm temperature dependent motional narrowing. Temperature dependent linewidth measurements showed a decrease in NMR linewidth of water protons with increasing temperature, coinciding closely with the increase in self-diffusion constant of water with temperature (except at low temperatures where, $\Delta\omega\tau_c \sim 1$, and motional narrowing is not effective). Figure 1 shows the linewidth of water protons for various temperatures and various concentrations of the superparamagnetic nanoparticles. The NMR linewidth scales linearly with the concentration of the particle, but the relative change in the linewidth of the protons is independent of the concentration, and scales with the diffusion constant of water [7].

At very low concentrations (≤ 65 ppm in our measurement) and at high temperatures, the natural linewidth of water due to imperfect shimming becomes comparable to the linewidth induced by the presence of nanoparticles, and then the decrease in linewidth with temperature deviates from the increase in diffusion constant. This is the lower limit for the concentration of the nanoparticles for linewidth to be a thermometer in our method – the linewidth induced by the particles should be significantly larger than the natural proton linewidth in the tissue. Empirically, we find this limit to be ~ 65 ppm. The results from the linewidth ($\Delta v$) measurements show that a $T_2^*$ contrast MRI imaging is a diffusion contrast MRI imaging in the presence of superparamagnetic nanoparticles as long as shimming imperfections do not dominate.

When $T_{Echo} \gg \tau_c = \frac{d^2}{4D}$, in the absence of external field inhomogeneities, $T_2$ is equivalent to $T_2^*$ [21], and therefore, a $T_2$ contrast imaging also is a diffusion contrast imaging. To establish the equivalence of $T_2$ and $T_2^*$ in the presence of the nanoparticles, $T_2^*$ and $T_2$ measurements ($T_{Echo}$ = 0.1 ms) were taken for 3 different concentrations of nanoparticles in water, and the results are included in Fig 2. The results show $T_2$ and $T_2^*$ are equivalent as if the linewidth induced by the nanoparticle is much greater than the natural linewidth of water due to improper shimming. At very low concentrations of nanoparticles and at high temperatures (65 ppm in NMR and 10 ppm in MRI), $T_2$ still remains a good thermometer, showing the same dependence to the diffusion constant, while $T_2^*$ is overwhelmed by shimming imperfections.

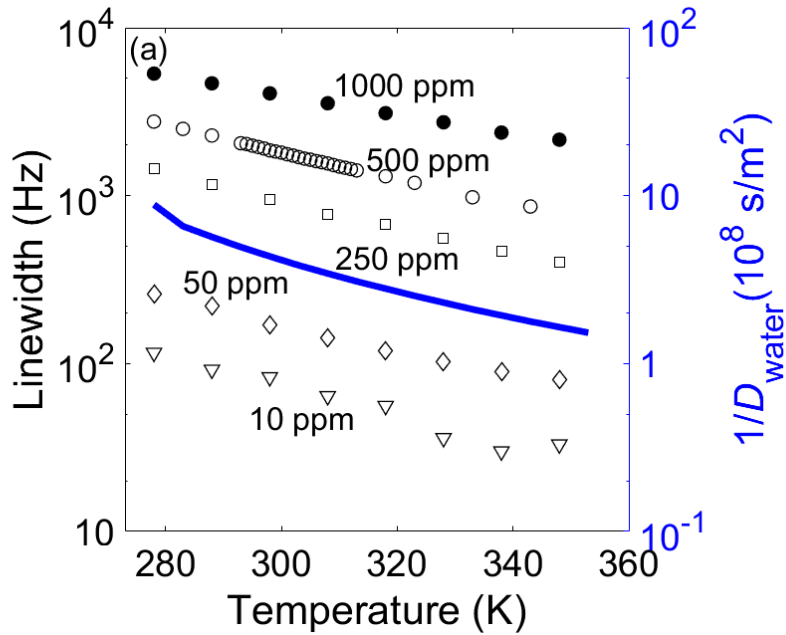

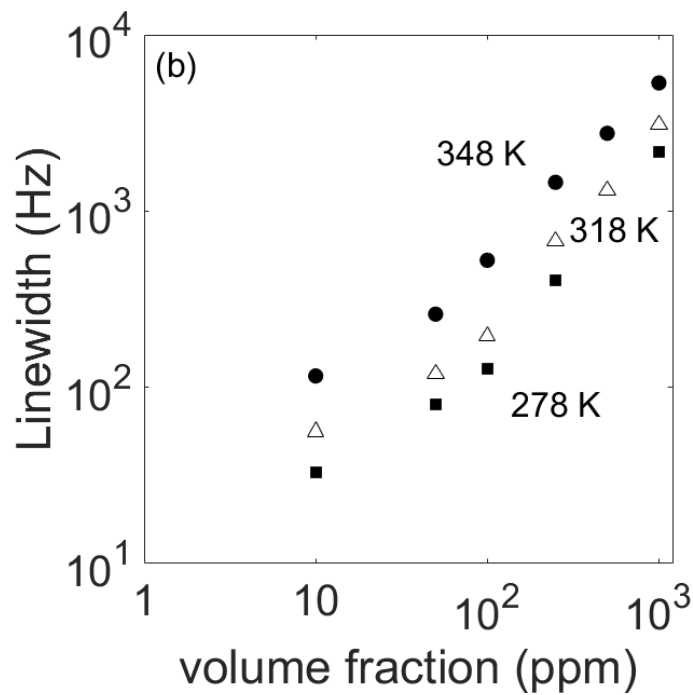

Fig1. (a) The temperature dependence of NMR linewidths (and consequently $T_2^*$) of protons in water in the presence of SPIONs and its comparison to the temperature dependence of the reciprocal of the diffusion coefficient of proton in water (blue solid line). The linewidths show the same temperature dependence as the reciprocal of the diffusion constant, until the concentration of the nanoparticles is low and the linewidth due to improper shimming becomes significant (at around 50 ppm nanoparticles by volume). (b) Comparison of NMR linewidth of protons in water with the volume fraction of SPOINs at 278 K (black squares), 318 K (open black triangles) and 348 K (black circles). The linewidths scale linearly with the volume fraction at all temperatures except for low volume fraction where shimming imperfections contribute significantly to the linewidth.

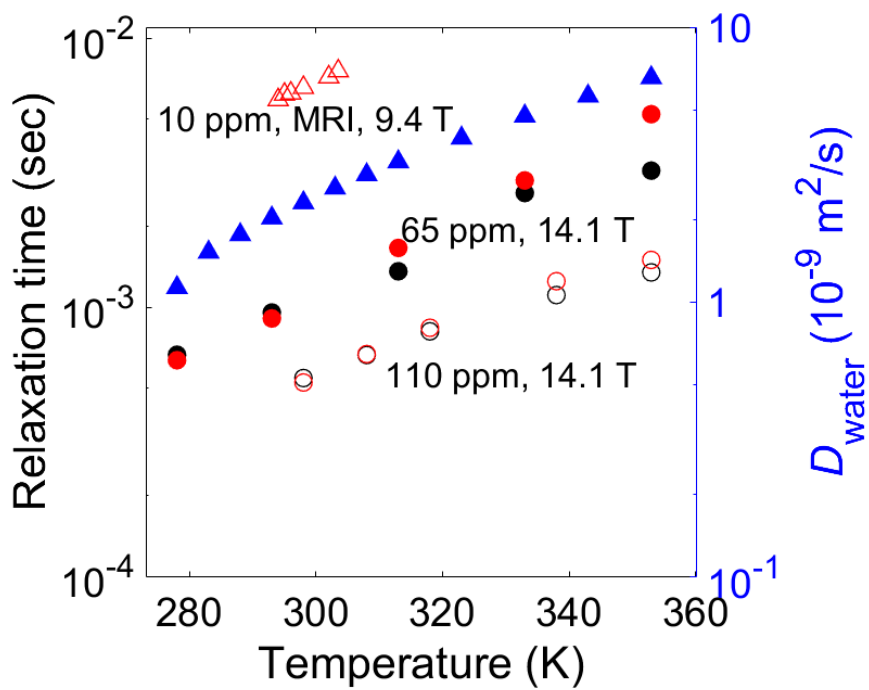

Fig 2. Temperature dependence of relaxation rates $1/T_2$ (red) and $1/T_2^*$ (black) at SPION volume fractions of 110 ppm (open circles), 65 ppm (filled circles) and 10 ppm (open triangles). The temperature dependence of the relaxation rates is compared with the temperature dependence of the reciprocal of the diffusion constant in water (blue solid line). The measurements of relaxation rates for SPION volume fractions of 110 ppm and 65 ppm were carried out in a Unity/Inova 14.1 T NMR spectrometer while the measurements for the volume fraction of 10 ppm was carried out in Bruker BioSpec 9.4 T preclinical MRI scanner. For lower volume fractions (< 65 ppm) and high temperature, $T_2^*$ has a significant contribution from field inhomogeneities due to imperfect shimming while $T_2$ is not affected and remains a good thermometer.

**$T_2$ mapping from MRI**

The applicability of the method was demonstrated in pure water (iron oxide nanoparticles present in 10 ppm by volume) on a 9.4 T Bruker BioSpec precilinical MRI system. Isothermal $T_2$ mapping were performed using multispin multiecho (MSME) sequence [28] at various temperatures in the range 21° C - 31° C. The isothermal $T_2$ measurements showed a standard deviation in $T_2$ of about 0.7 % for a specific temperature. Given the sensitivity of ~ 4 at room temperature, this variation in $T_2$ corresponds to a temperature resolution of approximately 0.5 K.

Figure 3 shows the increase in average $T_2$ with increasing temperature (specific values of $T_2$ vs temperature are included in Fig 2). As expected from the NMR measurements (see Fig 2), the increase in $T_2$ corresponds to the increase in the self-diffusion constant of water, even for a concentration as small as 10 ppm in pure water.

If no external field inhomogeneities are present, a $T_2^*$ mapping should be equivalent to a temperature mapping. However, due to the small concentration of nanoparticles used in the MRI measurement, and due to the inability to perform a good shimming due to the particles, the external field inhomogeneities were significant enough to cause a larger variation of $T_2^*$. Therefore, $T_2^*$ mapping had a significantly worse temperature resolution. This result emphasizes the fact that our method utilizing the motional narrowing on $T_2^*$ is a more reliable thermometry compared to $T_2^*$ methods utilizing the temperature dependent magnetization of contrast agents.

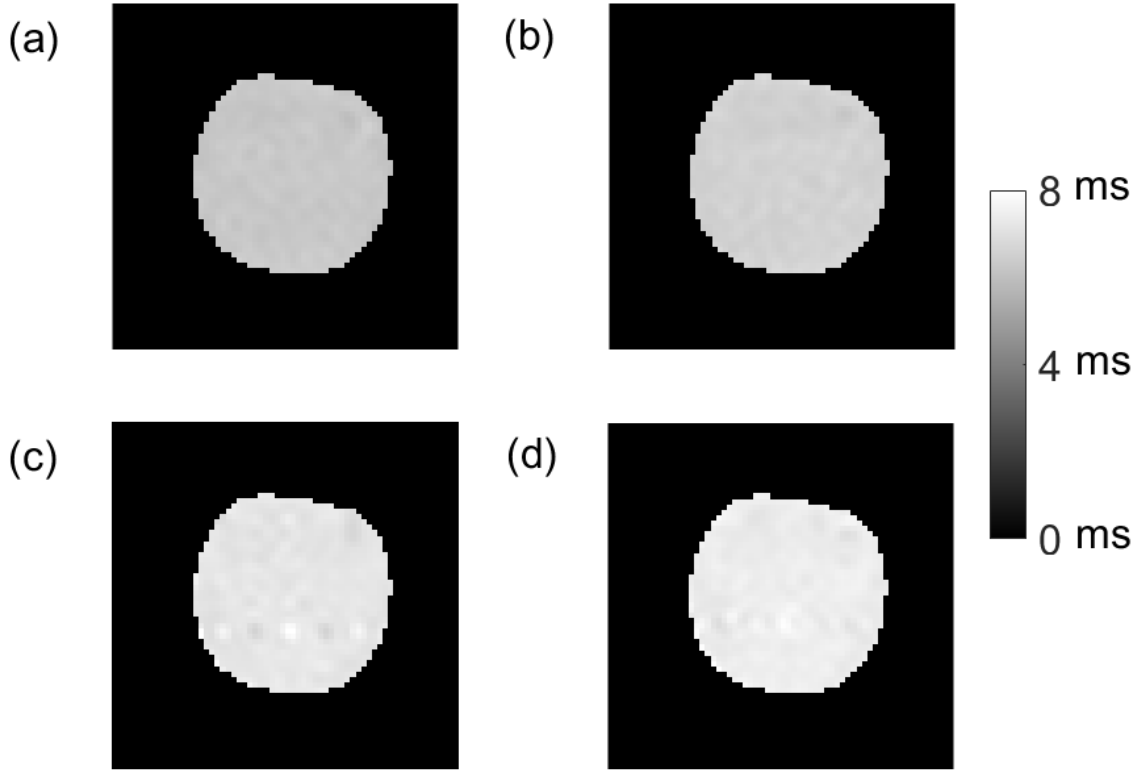

Fig 3. $T_2$ maps obtained from cross sectional (coronal) images of water in a cylindrical container at 23 °C (a), 25 °C (b), 29 °C (c) and 31 °C (d) in the presence of SPIONs (10 ppm by volume). The $T_2$ maps are obtained from 12 MSME images with equal echo spacing of 5 ms. The grey scale represents $T_2$ values with black corresponding to 0 ms and white corresponding to 8 ms. The average echo times determined from the $T_2$ maps are 6.25 ms, 6.57 ms, 7.19 ms and 7.42 ms respectively.

## V. DISCUSSION

Our results show, in the presence of superparamagnetic nanoparticles in the motional narrowing regime, a $T_2$ mapping is a diffusion mapping. However, since $T_2$ imaging does not require any interscan baseline image, it is not susceptible to motional artifacts unlike conventional diffusion imaging. Additionally, since $T_2$ mapping and especially $T_2$ contrast imaging is intrinsically faster than diffusion contrast imaging. Therefore, $T_2$ imaging provides a high-resolution thermometry in a relatively short scan time.

One requirement for this method to be applicable is that that the correlation time of protons diffusing past a SPION needs to be smaller than the reciprocal of frequency shift induced by the SPION: $\Delta\omega\tau_c \leq 1$, i.e., motional narrowing needs to take place. For the particles used in the experiment ($M_p = 3.0 \times 10^5$ A/m and $d = 13.5$ nm), $\Delta\omega \approx 3 \times 10^7$ s$^{-1}$, and in pure water $\tau_c = 2 \times 10^{-8}$ s at room temperature ($D = 2.5 \times 10^{-9}$ m$^2$/s). Therefore, the condition of motional narrowing is satisfied. This condition would also be marginally satisfied in the extracellular water in blood at body temperature where the diffusion constant is 40% smaller and consequently the correlation time is 40% larger for the same size of nanoparticles [10]. The condition is, however, not satisfied for intracellular water typically having a diffusion coefficient of ~ $8 \times 10^{-10}$ m$^2$/s [10].

We can define a critical diameter of the particle which satisfies the condition $\Delta\omega\tau_c = 1$. Since $\Delta\omega = \sqrt{\frac{4}{5}\frac{\gamma\mu_0 M_p}{3}}$ and $\tau_c = \frac{d^2}{4D}$, the critical diameter that just satisfies the motional narrowing condition, $\Delta\omega\tau_c = 1$, is given by: $d_c = \left(\sqrt{\frac{4}{5}\frac{\gamma\mu_0 M_p}{12D}}\right)^{-\frac{1}{2}}$. For nanoparticles with the magnetization used in our experiment, $d_c \approx 18.5$ nm in pure water, while for intracellular water where $D \approx 8 \times 10^{-10}$ m$^2$/s, $d_c \approx 10$ nm.

The effect on relaxation time when a fraction of particles has diameter greater than critical diameter is discussed in Supplemental Materials II [29].

It can be seen from the discussion in Supplemental Materials II that $T_2$ scales linearly with the diffusion constant for particles with diameters smaller than $d_c$ and inversely with the diffusion constant for particles with diameters larger than $d_c$, it is important that the fraction of particles above the critical diameter is not significant for our method to work.

The other two requirements for our method to be generally applicable to fluids are that the superparamagnetic particles significantly spoil the $T_2$ (i.e. $T_{2,S} \ll T_{2,0}$) and the echo pulses are not able to recover the relaxation due to the SPIONs ( i.e. $\tau_c \ll T_{echo} \sim T_{2,S}$). Even for fluids with $T_2$ as small as 50 ms (which is less than $T_2$ values of protons in most tissues [32]), a $T_{2,S}$ of ~10 ms can be obtained with a SPION volume fraction of ~10 ppm, which is a safe quantity for medical applications [19]. For thermal management fluids, there is no restriction on volume fraction due to toxicity. Therefore, the first of these requirements is easily satisfied. Even for a very small $T_{echo}$ of 10 ms, which would be required for fluids with a small $T_{2,S}$ of ~10 ms. If the diffusion constant is not very small, the correlation time remains much smaller than 10 ms, and the second condition is also easily satisfied. Therefore, with a careful selection of SPOIN size and volume fraction, our method can be generally applied to water in tissues and blood as well as in thermal management fluids.

In vivo application in biological samples, however, will be challenging due to different values and temperature dependence of the self-diffusion constant of water in different tissues. This is, however, a general challenge in MR thermometry [1] and is not a problem unique to our method.

Application to immersion cooling or heat sink applications is restricted to length scales of a few hundred microns to a few centimeters. The upper limit on the length scale is restricted by the availability of MR scanners that can accommodate large scale heat systems (e.g., cooling of data centers [33]) where immersion cooling is applied. The lower limit of length scale is determined by the signal-to-noise ratio from a voxel. Typically, it is limited to voxel of $(100\ \mu\text{m})^3$. At smaller length scales paramagnetic effects and RF absorption due to metals [34] used in devices would also make this method difficult to apply.

Finally, we note that the sensitivity of the thermometry can further be improved by combining motional narrowing with contrast agents that show a significant change in magnetization at temperatures of interest. For example, gadolinium silicide contrast agents show a sharp change in magnetization at approximately 30 °C [35]. Using nanoparticles of gadolinium silicide or similar contrast agents could allow utilizing both motional narrowing and the temperature dependence of magnetization for biological thermometry with higher sensitivity.

## ACKNOWLEDGEMENTS

The authors thank Dr. Shreyan Majumdar of the Biomedical Imaging Center, Beckman Institute, University of Illinois for assistance with the MRI measurements and Dr. Lingyang Zhu of the School of Chemical Sciences, University of Illinois for assistance with the NMR measurements. This research was funded by Semiconductor Research Corporation (Task ID: 3044.0001). This work was conducted in part at the Biomedical Imaging Center of the Beckman Institute for Advanced Science and Technology at the University of Illinois Urbana-Champaign (UIUC-BI-BIC). The Quantum Design MPMS3 SQUID Magnetometer used in the experiment was partially funded through Illinois MRSEC NSF Award Number DMR-1720633.

## REFERENCES

[1]   V. Rieke and K. B. Pauly, *MR Thermometry*, J. Magn. Reson. Imaging **27**, 376 (2008).

[2]   C. Stefanadis, C. Chrysohoou, E. Paraskevas, D. B. Panagiotakos, D. Xynopoulos, D. Dimitroulopoulos, K. Petraki, C. Papadimitriou, K. Karoutsos, C. Pitsavos, and P.V. Toutouzas, *Thermal Heterogeneity Constitutes a Marker for the Detection of Malignant Gastric Lesions in Vivo*, J. Clin. Gastroenterol. **36**, 215 (2003).

[3]   M. Suresh Patil, J. H. Seo, and M. Y. Lee, *A Novel Dielectric Fluid Immersion Cooling Technology for Li-Ion Battery Thermal Management*, Energy Convers. Manag. **229**, 113715 (2021).

[4]   D. Bihan, J. Delaanoy, and R. Levin, *Temperature Mapping with MR Imaging of Molecular Diffusion : Application to Hyperthermia*, Ther. Radiol. **171**, (1989).

[5]   F. Bertsch, J. Mattner, M. K. Stehling, U. Müller-Lisse, M. Peller, R. Loeffler, J. Weber, K. Meßmer, W. Wilmanns, R. Issels, and M. Reiser, *Non-Invasive Temperature Mapping Using MRI: Comparison of Two Methods Based on Chemical Shift and T1-Relaxation*, Magn. Reson. Imaging **16**, 393 (1998).

[6]   Y. Ishihara, A. Calderon, H. Watanabe, K. Okamoto, Y. Suzuki, K. Kuroda, and Y. Suzuki, *A Precise and Fast Temperature Mapping Using Water Proton Chemical Shift*, Magn. Reson. Med. **34**, 814 (1995).

[7]   M. Holz, S. R. Heil, and A. Sacco, *Temperature-Dependent Self-Diffusion Coefficients of Water and Six Selected Molecular Liquids for Calibration in Accurate 1H NMR PFG Measurements*, Phys. Chem. Chem. Phys. **2**, 4740 (2000).

[8]   D. C. Chang, H. E. Rorschach, B. L. Nicholls, and C. F. Hazlewood, *Implications of Diffusion Coefficient Measurements for the Structure of Cellular Water*, Ann. New York

Acad. Sci. **204**, 434 (1973).

[9] J. G. Li, G. J. Stanisz, and R. M. Henkelman, *Integrated Analysis of Diffusion and Relaxation of Water in Blood*, Magn. Reson. Med. **40**, 79 (1998).

[10] I. Åslund and D. Topgaard, *Determination of the Self-Diffusion Coefficient of Intracellular Water Using PGSE NMR with Variable Gradient Pulse Length*, J. Magn. Reson. **201**, 250 (2009).

[11] M. Hermansson, E. Johansson, and M. Jansson, *Self-Diffusion of Oil in Lubricating Greases by NMR*, J. Synth. Lubr. **13**, 279 (1996).

[12] M. H. Rausch, L. Kretschmer, S. Will, A. Leipertz, and A. P. Fröba, *Density, Surface Tension, and Kinematic Viscosity of Hydrofluoroethers HFE-7000, HFE-7100, HFE-7200, HFE-7300, and HFE-7500*, J. Chem. Eng. Data **60**, 3759 (2015).

[13] I. R. Jeon, J. G. Park, C. R. Haney, and T. D. Harris, *Spin Crossover Iron(Ii) Complexes as PARACEST MRI Thermometers*, Chem. Sci. **5**, 2461 (2014).

[14] J. H. Hankiewicz, Z. Celinski, K. F. Stupic, N. R. Anderson, and R. E. Camley, *Ferromagnetic Particles as Magnetic Resonance Imaging Temperature Sensors*, Nat. Commun. **7**, 1 (2016).

[15] Wahajuddin and S. Arora, *Superparamagnetic Iron Oxide Nanoparticles: Magnetic Nanoplatforms as Drug Carriers*, Int. J. Nanomedicine **7**, 3445 (2012).

[16] Y.-X. J. Wang, S. M. Hussain, and G. P. Krestin, *Superparamagnetic Iron Oxide Contrast Agents: Physicochemical Characteristics and Applications in MR Imaging*, Eur. Radiol **11**, 2319 (2001).

[17] B. Gleich and J. Weizenecker, *Tomographic Imaging Using the Nonlinear Response of Magnetic Particles*, Nature **435**, 1214 (2005).

[18] S. Laurent, D. Forge, M. Port, A. Roch, C. Robic, L. Vander Elst, and R. N. Muller, *Magnetic Iron Oxide Nanoparticles: Synthesis, Stabilization, Vectorization, Physicochemical Characterizations and Biological Applications*, Chem. Rev. **108**, 2064 (2008).

[19] J. M. Poller, J. Zaloga, E. Schreiber, H. Unterweger, C. Janko, P. Radon, D. Eberbeck, L. Trahms, C. Alexiou, and R. P. Friedrich, *Selection of Potential Iron Oxide Nanoparticles for Breast Cancer Treatment Based on in Vitro Cytotoxicity and Cellular Uptake*, Int. J. Nanomedicine **12**, 3207 (2017).

[20] M. Gueron, *Nuclear Relaxation in Macromolecules by Paramagnetic Ions: A Novel Mechanism*, J. Magn. Reson. **19**, 58 (1975).

[21] R. Van Roosbroeck, W. Van Roy, T. Stakenborg, J. Trekker, A. D'Hollander, T. Dresselaers, U. Himmelreich, J. Lammertyn, and L. Lagae, *Synthetic Antiferromagnetic Nanoparticles as Potential Contrast Agents in MRI*, ACS Nano **8**, 2269 (2014).

[22] S. Tong, S. Hou, Z. Zheng, J. Zhou, and G. Bao, *Coating Optimization of Superparamagnetic Iron Oxide Nanoparticles for High T2 Relaxivity*, Nano Lett. **10**, 4607 (2010).

[23] R. Brooks, F. Moiny, and P. Gillis, *On T2-Shortening by Weakly Magnetized Particles: The Chemical Exchange Model*, Magn. Reson. Imaging **45**, (2001).

[24] J. Wang, W. Wu, F. Zhao, and G. M. Zhao, *Curie Temperature Reduction in SiO2 -Coated*

*Ultrafine Fe 3 O4 Nanoparticles: Quantitative Agreement with a Finite-Size Scaling Law*, Appl. Phys. Lett. **98**, 1 (2011).

[25] C. Nayek, K. Manna, G. Bhattacharjee, P. Murugavel, and I. Obaidat, *Investigating Size- and Temperature-Dependent Coercivity and Saturation Magnetization in PEG Coated Fe3O4 Nanoparticles*, Magnetochemistry **3**, (2017).

[26] H. Y. Carr and E. M. Purcell, *Effects of Diffusion on Free Precession in Nuclear Magnetic Resonance Experiments*, Phys. Rev. **94**, 630 (1954).

[27] D. S. Raiford, C. L. Fisk, and E. D. Becker, *Calibration of Methanol and Ethylene Glycol Nuclear Magnetic Resonance Thermometers Determination of Cobalt by Lophine Chemiluminescence*, **51**, 2050 (1979).

[28] Y. Fatemi, H. Danyali, M. S. Helfroush, and H. Amiri, *Fast T2 Mapping Using Multi-Echo Spin-Echo MRI: A Linear Order Approach*, Magn. Reson. Med. **84**, 2815 (2020).

[29] See Supplemental Material at [ ] for SQUID magnetometry data and calculation of the relaxation rate for fraction of particles with diameters larger than the critical diameter.

[30] J. Qin, S. Laurent, Y. S. Jo, A. Roch, M. Mikhaylova, Z. M. Bhujwalla, R. N. Müller, and M. Muhammed, *A High-Performance Magnetic Resonance Imaging T2 Contrast Agent*, Adv. Mater. **19**, 1874 (2007).

[31] B. Friedrich, J. Auger, S. Dutz, I. Cicha, E. Schreiber, J. Band, A. R. Boccacccini, G. Krönke, C. Alexiou, and R. Tietze, *Hydroxyapatite-Coated SPIONs and Their Influence on Cytokine Release*, (2021).

[32] G. J. Stanisz, E. E. Odrobina, J. Pun, M. Escaravage, S. J. Graham, M. J. Bronskill, and R.

M. Henkelman, *T1, T2 Relaxation and Magnetization Transfer in Tissue at 3T*, Magn. Reson. Med. **54**, 507 (2005).

[33] I. W. Kuncoro, N. A. Pambudi, M. K. Biddinika, I. Widiastuti, M. Hijriawan, and K. M. Wibowo, *Immersion Cooling as the next Technology for Data Center Cooling: A Review*, J. Phys. Conf. Ser. **1402**, 2 (2019).

[34] S. Chandrashekar, N. M. Trease, H. J. Chang, L. S. Du, C. P. Grey, and A. Jerschow, *7Li MRI of Li Batteries Reveals Location of Microstructural Lithium*, Nat. Mater. **11**, 311 (2012).

[35] M. Nauman, M. H. Alnasir, M. A. Hamayun, Y. Wang, M. Shatruk, and S. Manzoor, *Size-Dependent Magnetic and Magnetothermal Properties of Gadolinium Silicide Nanoparticles*, RSC Adv. **10**, 28383 (2020).

[36] Q. L. Vuong, P. Gillis, and Y. Gossuin, *Monte Carlo Simulation and Theory of Proton NMR Transverse Relaxation Induced by Aggregation of Magnetic Particles Used as MRI Contrast Agents*, J. Magn. Reson. **212**, 139 (2011).